\documentclass[twocolumn,tighten]{aastex63}

\usepackage{times,natbib,graphicx,amsmath,multirow,xspace}
\usepackage{xcolor}
\usepackage{lineno}
\usepackage{rotating}
\usepackage{longtable}

\newcommand{\chandra}{Chandra\xspace}

\begin{document}

\title{The Feasibility of Using Fe {\sc XXIII} Metastable Transitions as a Density Diagnostic for LMXB Disk Winds}

\correspondingauthor{D. L. Moutard}
\email{moutard@umich.edu}


\author[0000-0003-1463-8702]{~D.~L.~Moutard}
\author[0000-0002-5466-3817]{~L.~R.~Corrales}
\affiliation{Department of Astronomy, University of Michigan, 1085 S. University, Ann Arbor, MI 48109, USA}
\author[0000-0002-6797-2539]{~R.~Tomaru}
\affiliation{Department of Earth and Space Science, Osaka University, 1-1 Machikaneyama-cho, Toyonaka-shi, Osaka 560-0043, Japan}
\author[0000-0002-1065-7239]{~C.~Done}
\affiliation{Centre for Extragalactic Astronomy, Department of Physics, Durham University, South Road, Durham DH1 3LE, UK}
\author[0000-0002-8247-786X]{~J.~Neilsen}
\affiliation{Department of Physics, Villanova University, 800 Lancaster Avenue, Villanova, PA 19085, USA}
\author[0000-0001-9735-4873]{~E.~Behar}
\affiliation{Physics Department, Technion, Haifa 32000, Israel}
\author[0000-0001-8470-749X]{~E.~Costantini}
\affiliation{SRON Netherlands Institute for Space Research, Niels Bohrweg 4, 2333 CA Leiden, the Netherlands}
\affiliation{Anton Pannekoek Institute for Astronomy, University of Amsterdam, Science Park 904, NL-1098 XH Amsterdam, the Netherlands}
\author[0000-0001-7796-4279]{~M.~D\'iaz-Trigo}
\affiliation{ESO, Karl-Schwarzschild-Strasse 2, 85748, Garching bei München, Germany}
\author[0000-0003-4808-893X]{~S.~Yamada}
\affiliation{Department of Physics, Rikkyo University, 3-34-1 Nishi Ikebukuro, Toshima-ku, Tokyo 171-8501, Japan}

\begin{abstract}
    Low mass X-ray binaries (LMXBs) occasionally show signs of outflowing material from the accretion disk. Studying these outflows can inform the understanding of the geometry of the systems, as well as the dynamics and energetics of accretion. One key variable for determining the location of these disk winds is the density of the outflowing material. In this paper we explore a density diagnostic based upon the absorption of ionizing photons by density-sensitive metastable states of Fe {\sc XXIII}. This can yield a blue shifted complex of absorption features in the region of $6.61-6.64$ keV. We use the photoionization code {\sc pion} to test how varying the ionizing spectrum affects the detectability and interpretation of these features. We base these ionizing spectral energy distributions on GX~13$+$1 to represent a bright thermally dominated spectrum; 4U 1735$-$44 representing a harder, fainter LMXB spectrum; and MAXI J1820$+$070 representing a black hole LMXB spectrum completely dominated by Comptonized emission. For each of these, we find that the regime where Fe {\sc XXIII} can be used as a density diagnostic is with an ionization parameter $\log{(\xi/{\rm erg~cm~s^{-1}})}\sim2-3$  and an outflow density $\log{(n_H/{\rm cm^{-3})}}\gtrsim14$. The typical range of ionization parameters for LMXBs indicates that this technique is more feasibly achieved with BH LMXBs than their NS counterparts. 
\end{abstract}

\section{Introduction}\label{sec:intro}

Low-mass X-ray binaries (LMXBs) are sources which contain a compact object such as a neutron star (NS) or black hole (BH) accreting material from a companion star of less than a few $M_\odot$ \citep{bahramian23}. LMXBs differ from their high-mass counterparts by accreting via a geometrically thin disk \citep{shakura73} rather than via a stellar wind from the companion \citep{fortin23}. This type of thin disk accretion is a useful analog for other disk-accretion systems, such as active galactic nuclei (AGN).  As such, LMXBs can be used as a local laboratory for studying more distant systems. The accretion process can generate X-rays from various different components and regions. These include the accretion disk that is formed by the in-falling stream of material and a corona comprised of hot electrons near the compact object which inverse Compton scatter seed photons from the disk. If the central source is a NS, there may also be an additional soft X-ray blackbody from the NS surface itself or a boundary layer where accreted material spreads over the surface \citep{syunyaev91,popham01}. 

LMXBs can be found in various states, depending on the central accreting object and the rate of accretion from the companion star. In the broadest sense, these are typically hard and soft states, usually associated with low and high luminosity respectively. The low-hard state is dominated by the Comptonized emission with relatively low flux due to a low mass accretion rate, and the high-soft state is typically dominated by disk emission, with a higher mass accretion rate leading to a higher luminosity \citep{homan05}. NS LMXBs are often divided by the shape of their track on a color-color or color-magnitude diagram. These are described as either Z or atoll sources, with Z sources tracing out three distinct connected branches in the diagram. Atoll sources on the other hand typically exist in the ``island" state (disconnected regions in the low-hard region of the color intensity diagram) or the ``banana" state (a vaguely banana shaped track in the high-soft region of the diagram \citealt{church14}). Atoll sources are fainter than their Z-source counterparts, and they also tend to differ in timing properties \citep{vanderklis89}.

Accreting systems can occasionally display signs of outflowing material. This is typically in the form of blueshifted absorption features seen in the X-ray spectrum, indicative of a disk wind along the line of sight. In LMXBs this is thought to be the result of thermal heating of the disk, a process which is more effective further from the center of the disk and therefore results in slower outflows \citep{begelman83, tomaru23}. Alternatively, winds can be launched more uniformly by magnetic field lines permeating the disk \citep{konigl94, miller08, chakravorty23}. These winds are typically variable, changing alongside the accretion properties and spectral state of the system \citep{diaztrigo12, petrucci21}. For recent overviews of disk wind studies, see \cite{diaztrigo16} and \cite{parra24}. 

Due to the high energy of the illuminating spectrum of LMXBs, the features seen in these disk-wind signatures arise from highly ionized material. This ionization is often described using a value called the ionization parameter ($\xi$), defined as
\begin{equation} \label{eq:xi}
    \xi = \frac{L}{n_Hr^2}
\end{equation}
 where $L$ is the $1-1000$ Ryd luminosity of the ionizing continuum, $n_H$ is the Hydrogen number density of the ionized material, and $r$ is the distance from the ionizing source to the ionized material \citep{tarter69}. While the values of $L$ and $\xi$ can be measured, the degeneracy between $n_H$ and $r$ is an obstacle to measuring outflow parameters. Understanding the location of the disk wind, and therefore constraining the geometry of accretion, is difficult without first understanding the density of the wind itself. 

One method of measuring density is to use specific density-sensitive metastable absorption lines. Metastable states can persist for a long time, and can provide insight into the density of the wind, due to the fact their occupation is driven largely by collisional interactions. This method has been used in the study of active galactic nucleus (AGN) disk winds. By searching for density-sensitive transitions of O {\sc v}, \cite{kaastra04} are able to place estimates on the density of the wind of the AGN Mrk 279 with \chandra.  Using the photoionization model {\sc PION} \citep{miller15, mehdipour16} within the spectral modeling software {\sc SPEX} \citep{kaastra96}, \cite{mao17} utilizes the AGN spectral energy distribution from \cite{mehdipour15a} and report many theoretical measurable density-sensitive absorption features. These features, when compared to the ground state population of the ion, can be used to determine the density of the outflow. They apply this concept to the AGN NGC 5548, and yield a range of results consistent with other measured values of the source.  While these studies are useful for understanding the disk winds of AGN, it is important to note that the ionizing SED of a LMXB is distinct from that of an AGN; AGN tend to lack the thermal contribution from the surface or boundary layer of the NS in the case of NS-LMXBs, and are typically higher luminosity. 

An example of similar density diagnostics conducted for LMXBs can be seen in \cite{miller08}, who observe the transiently accreting stellar mass BH GRO J1655-40. Using L-shell transitions of Fe {\sc XXII} near 1 keV, they constrain the density of the outflow. In this paper we instead focus on density sensitive K-shell transitions of Fe {\sc XXIII}; details of these states and their transitions can be seen in Table \ref{tab:lines}.  It should be noted that the $^3P_1$ state is not truly metastable, as it decays to the ground state much more rapidly than the other states. This means that it does not appreciably contribute to to the metastable populations until density is very high. In Table \ref{tab:lines}, we also include a handful of other nearby transitions from these metastable states, retrieved from \cite{kramida24}. These are not associated with an oscillator strength, making it unclear how strong these features would appear. However, with future atomic database improvements, these could represent an additional means of disentangling transitions from metastable states.

\begin{table}[h!]

\caption{Fe XXIII and Fe XXIV Absorption Lines}
\centering

\begin{tabular}{c|c|c|c}
\hline\hline
Lower & Upper & $E_0$ (keV) & $f$\\
\hline
\multicolumn{4}{c}{Fe {\sc XXIII}}\\ \hline

2s$^2$ ($^1$S$_0$)$^a$ & 1s2s$^2$2p ($^1$P$_1$) &6.6288$^\dagger$ & 0.69  \\
\hline
2s2p ($^3$P$_0$)$^b$ & 1s2s2p$^2$ ($^3$P$_1$) &6.6194$^\dagger$ & 0.61 \\\hline
\multirow{4}{5em}{2s2p ($^3$P$_1$)$^{*}$}& 1s2s2p$^2$ ($^3$D$_2$) &6.6158 & 0.29 \\
& 1s($^2$S)2s2p2($^4$P) ($^3$P$_0$) &6.6118 & -- \\
& 1s($^2$S)2s2p2($^2$D) ($^3$D$_2$) &6.6173 & -- \\
& 1s($^2$S)2s2p2($^2$D) ($^3$D$_1$) &6.6233 & -- \\\hline
\multirow{2}{5em}{2s2p ($^3$P$_2$)$^b$ }& 1s2s2p$^2$ ($^3$P$_2$) &6.6158 & 0.26 \\
&1s($^2$S)2s2p2($^2$D) ($^3$D$_1$) &6.6119 &--\\

\hline
\multicolumn{4}{c}{Fe {\sc XXIV}} \\ \hline

1s$^2$2s ($^2$S$_{1/2}$)$^{a}$ & 1s($^2$S)2s2p($^3$P) ($^4$P$_{3/2}$) & 6.6167 & 0.017\\
\hline

\end{tabular}
\label{tab:lines}
\newline

{\it Note:} The ``Lower" column displays the electron configuration of the ground or metastable states, with the spectroscopic notation in parentheses. ``Upper" is the same but for the excited state from those levels. $E_0$ is the rest-frame energy of each transition, and $f$ is the oscillator strength. All values for Fe {\sc XXIII} in this table come directly from \cite{mao17}, unless noted with $^\dagger$, which indicates the value comes from \cite{steinbrugge22}. These values are deemed more accurate since they arise from experimental measurements and the resulting uncertainties are quite small (typically $\lesssim 0.1$eV). The values for Fe {\sc XXIV} come from \cite{mehdipour15a}. The first four transitions listed are for Fe {\sc XXIII}, and the last is for a contaminatingmFe {\sc XXIV} feature. Fe XXIII lines without reported $f$ values come from \cite{kramida24}. These lines do not represent every possible transition from the metastable states. 
$^a$ Ground states transitions.
$^b$ Metastable state transitions.
$^*$ Not a ``true" metastable state. This state decays to the ground state rapidly until very high densities. 

\end{table}

The recently launched X-Ray Imaging and Spectroscopy Mission (XRISM) provides a new avenue for conducting these studies. XRISM hosts the microcalorimeter instrument {\it Resolve}, which is capable of X-ray spectroscopy with an energy resolution of 4.5 eV \citep{xrism20}. This fine resolution presents the opportunity to search for narrow density sensitive absorption features in LMXB outflows. The effective energy band of XRISM {\it Resolve} ranges from 1.7$-$12 keV, with a maximum effective area near 6 keV.

The energy range of XRISM {\it Resolve} precludes Fe {\sc XXII} from being used as a density diagnostic, motivating our focus instead on the metastable density sensitive transitions of Fe~{\sc XXIII} for various LMXB ionizing SEDs. We use {\sc PION} with these SEDs to produce population curves at various densities, providing a reference for determining the density of the outflow. The present work expands on previous density diagnostics using Fe {\sc XXIII} by applying it to three different ionizing SEDs, each with two different column densities. These SEDs also extend into the EUV, probing effects of UV photoexcitation. In Section \ref{sec:methods} we describe three general ionizing SEDs and the LMXBs upon which they were based and use {\sc pion} to generate population curves for the features of interest. In Section \ref{sec:results} we discuss the implications of these findings and challenges that arise when using this method, and in Section \ref{sec:conc} we summarize our conclusions.

\section{Modeling and Methods}\label{sec:methods}
In the following section, we describe the process whereby we generate three unique SEDs in order to test the feasibility of using Fe~{\sc XXIII} as a density diagnostic for the winds of LMXBs. This will provide insight into the behavior of these density sensitive features not only as the density varies, but also for different sources, which provide a different ionizing continuum.

\subsection{Sources used for SED generation}\label{subsec:sources}
We conduct our analysis using three distinct ionizing SEDs, drawn from archival Chandra and XMM-Newton data, as well as upcoming XRISM results: a bright, thermally dominated LMXB spectrum (GX~13$+$1); a fainter LMXB with contributions from both thermal and non-thermal components (4U 1735$-$44); and an LMXB dominated entirely by non-thermal power law emission {(MAXI J1820$+$070). The spectrum of each source is fit using {\sc spex} v.3.08.01., then the fit values are extended from 0.01 to 30 keV. This range is chosen to avoid complications from the Compton hump in the hard X-rays, which can't be properly modeled using Chandra or XMM-Newton. We avoid this by cutting the spectra at a lower energy, while still accounting for the bulk of the flux in the region of the features of interest. Beyond 30 keV, the ionization cross sections of the ions of interest is negligible, so this cut does not have an appreciable impact on the occupations. 

It should be noted that UV photoexcitation can have an appreciable impact on the populations of metastable states. While these SEDs do extend into the extreme UV, this band is not directly measured, and likely does not accurately capture the UV behavior. In general, assuming a higher UV flux serves to increase the population of the metastable states \citep{mitrani23}. For example, SED-2 in \cite{tomaru23} has a UV contribution that is $\sim1-2$ orders of magnitude higher than that of SED-1 in the same study, which corresponds to an enhancement of the metastable population of roughly 50\%. Similarly, X-ray photoexcitation from lower energy transitions (e.g. n=2-3) can impact the population of metastable states.

\begin{figure}[t!h]
    \centering
    \includegraphics[width=0.98\linewidth, trim = 0 0 0 0, clip]{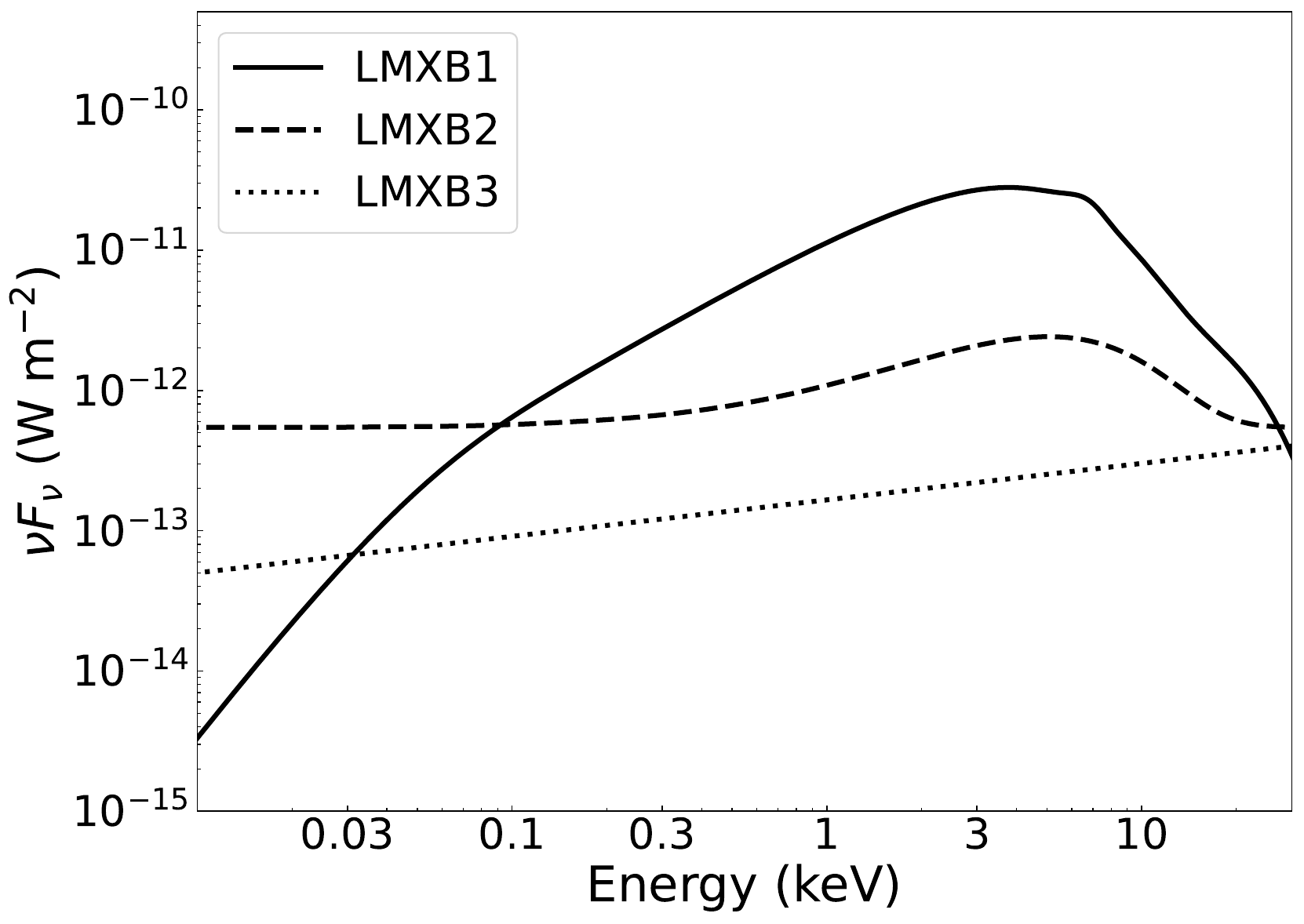}
    \caption{The three representative LMXB SEDs defined in the range from 0.01 to 30 keV. The SEDs with a power law component (LMXB2 and LMXB3) show more flux in the softest bands, but the LMXB1 SED has a higher overall luminosity. The flux excess in bands below $\sim$0.1 keV has little effect on the population of ions producing absorption features above 6.5 keV.}
    \label{fig:seds}
\end{figure}

\subsubsection{LMXB1: GX~13$+$1}
We generate an SED called LMXB1 based on the XRISM spectrum of GX~13$+$1 as reported in \cite{xrism25}, which we refer to hereafter as XC25. GX~13$+$1 is a NS LMXB \citep{fleishchman85} hosting a K5 III star located at a distance of $7\pm1$ kpc \citep{bandopadhyay99}, and displaying an orbital period of $\sim24$ days \citep{corbet10}. This source displays signatures of strong outflows, and as a result, it has been studied closely in the context of disk winds. The spectrum is dominated by soft blackbody components, representing the disk and emission from the NS or boundary layer surrounding it. This source is an example of a Z source NS LMXB \citep{kaddouh24}, named after the track it traces on a color-color diagram. However, in earlier studies it was uncertain whether GX~13$+$1 was in fact a Z source, or instead a bright example of a (typically faint) atoll source \citep{hasinger89, homan98}. While the source itself is representative of a Z-source LMXB, the XRISM observation captured it in a super-Eddington accretion state not seen before by Chandra or XMM-Newton (XC25). 

The SED is directly taken from the continuum model in XC25. The SPEX parameter values are adapted from the {\sc xspec} model fits reported in XC25, which utilizes different continuum parameter definitions. As such, the SED used in this work is an approximation of the fit reported in that paper, which represents a super-Eddington thermally dominated NS LMXB spectrum.

\subsubsection{LMXB2: 4U 1735$-$44}

We select 4U 1735$-$44 to use as the basis for an SED (hereafter LMXB2) to represent an atoll source LMXB. It is known to have a NS accretor \citep{mcclintock78, vanparadijs88}, and is located at a distance $5.6^{+2.1}_{-4}$ kpc \citep{arnason21} with a relatively short orbital period of $4.65\pm0.01$h \citep{corbet86}. Atoll sources, like Z sources, get their name from the track they follow on color-color diagrams. They are typically fainter than Z sources, due to a much lower mass accretion rate \citep{hasinger89}. 

The values used for this SED are based on a fit to a Chandra observation of 4U 1735$-$44 (OBSID = 704). This is representative of a NS LMXB which displays both thermal emission as well as hard power law emission representative of the corona \citep{ludlam20}. We set the powerlaw cutoff energy $E_0$ and change in powerlaw index $\Delta\Gamma$ to 150 keV and 3.0, respectively. This energy cutoff is beyond the range of the SED, and thus will not affect the overall shape.

\subsubsection{LMXB3: MAXI J1820$+$070}

We create an SED based on the somewhat recently discovered MAXI J1820$+$070 (hereafter LMXB3) to represent BH LMXBs in a hard state. MAXI J1820$+$070 is a LMXB with black hole mass measured to be $7.6\pm1.7$ M$_\odot$ located at a distance $3.06^{+1.54}_{-0.82}$ kpc \citep{tucker18, torres20, demarco21}. 

This SED is based off of a fit to an XMM observation of MAXI J1820$+$070 (OBSID = 844230301). Again, we set  E$_0=150$ keV and $\Delta\Gamma=3.0$. Although the XMM band is used to fit only ranges between $0.4-2.0$ keV, we find that the fit results are comparable to those found in previous studies of the hard state \citep{demarco21}.

\begin{table}[h!]
\centering

\caption{LMXB SED parameters}

\begin{tabular}{ll|ccc}
  & &LMXB1 &LMXB2 &LMXB3 \\

\hline
  \multirow{2}{*}{{\sc dbb}} &kT (keV) &3.08 &4.21 &---\\ 
  & norm$^a$  &1.85$\times10^{-5}$ &3.59$\times10^{-7}$ &---\\ 
\hline
  \multirow{2}{*}{{\sc bb}} &kT (keV) &3.3 &--- &---\\ 
  & norm$^a$  &3$\times10^{-6}$ &--- &---\\ 
\hline
  \multirow{2}{*}{{\sc pow}} &$\Gamma$ &--- &2.0 &1.74\\ 
  & norm$^b$  &--- &4.27$\times10^4$ &1.4$\times10^4$\\ 

\hline
  \multirow{3}{*}{{\sc gaus}} &E (keV) &6.4 &--- &---\\ 
  &FWHM (keV)  &1.91 &--- &---\\ 
  & norm$^c$  &1.4$\times10^4$ &--- &---\\ 
\hline
\end{tabular}
\medskip

{\it Notes:} $^a$normalization for both the blackbody and the disk blackbody is in units $10^{16}$ m$^2$. $^b$normalization for the power law is in units $10^{44}$ph s$^{-1}$ kev$^{-1}$ at 1 keV.  $^c$normalization for the gaussian is in units $10^{44}$ph s$^{-1}$. The SEDs are simulated from 0.01 to 30 keV to avoid complications due to unmodeled reflection components at high energies. For both SEDs with {\sc pow} components, the $E_0$ values are set to 150 keV, beyond the range of the simulated SED. 
\label{tab:seds}
\end{table}
\subsection{{\sc PION} Model Grid} \label{subsec:pion}
To generate SEDs based on the sources above, we begin with the model components described in Table \ref{tab:seds} and shown in Figure \ref{fig:seds} using {\sc SPEX}. We then apply {\sc pion} to each component using default parameters. The only parameter changed is the covering factor parameter {\sc omeg}, which we set to an arbitrarily small value of $10^{-9}$. Typically this factor is only non-zero when studying emission, but a value $>0$ is also required to conduct density diagnostics. We then vary $\log \xi$ from $2$ to $4$, calculating the concentrations of Be-like Fe {\sc XXIII} and Li-like Fe {\sc XXIV} at each step, and for each of the LMXB SEDs. The Hydrogen column density $N_{\rm H}$ is set to a value to match that of XC25 ($\log{(N_{\rm H}/\mathrm{cm}^{-2})} = 24.11$). The results of this are shown in Figure \ref{fig:conc}. 

\begin{figure}[t!h]
    \centering
    \includegraphics[width=0.98\linewidth, trim = 0 0 0 0, clip]{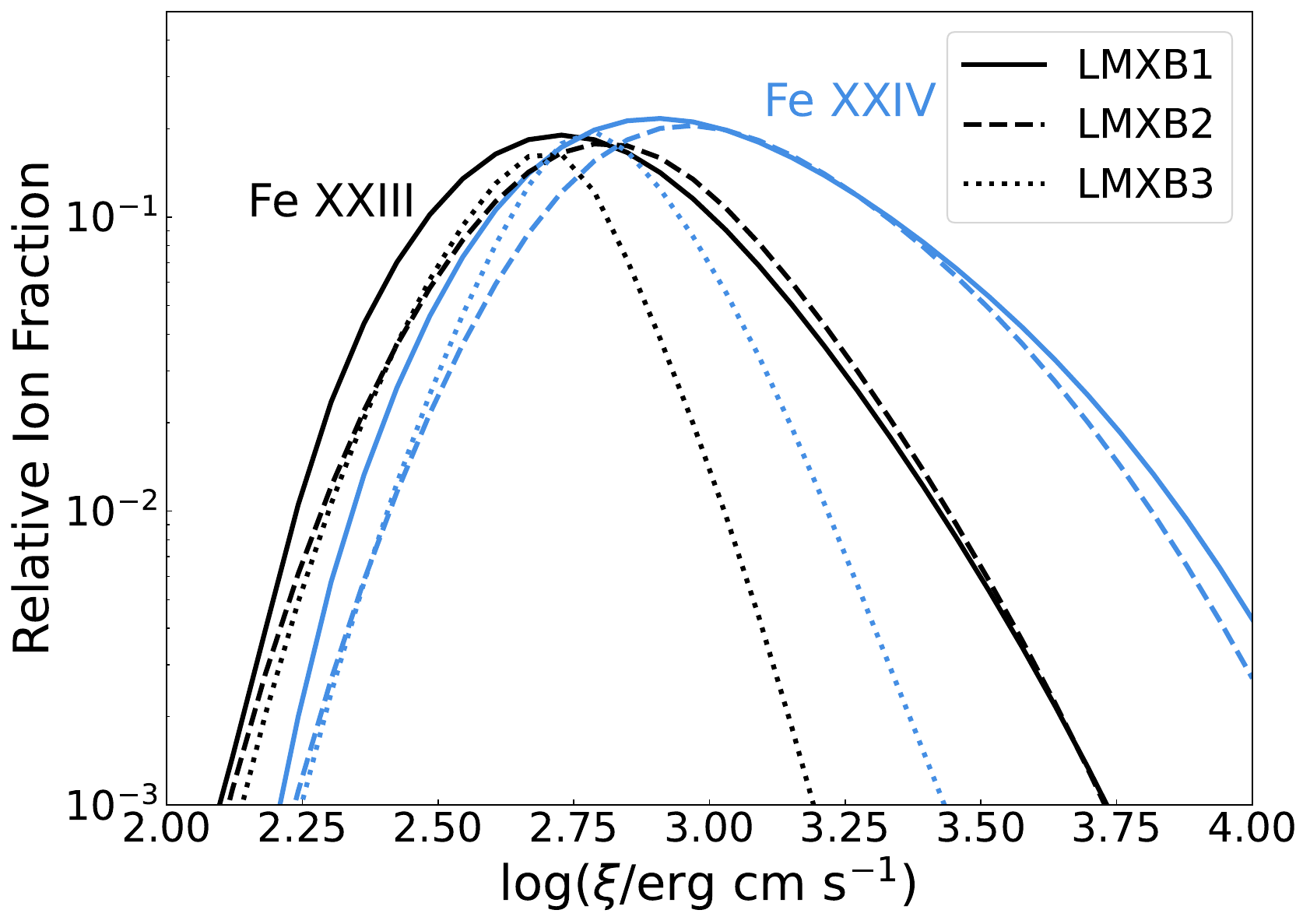}
    \caption{The relative ion fraction of Fe {\sc XXIII} and {\sc XXIV} for various ionizing LMXB SEDs. From this plot, it is clear that the nature of the SED has a strong effect on the concentrations of relevant ions.}
    \label{fig:conc}
\end{figure}

After calculating the ion concentrations at various ionization parameters, we look to see how different metastable absorption features vary with outflow density. We do this ion-by-ion, first fixing $\xi$ to a value which maximizes the concentration of the ion in question (e.g. we fix $\log{(\xi/\mathrm{erg~cm~s}^{-1})} = 2.73$  for Fe {\sc XXIII}). Then, we increase the wind density $n_H$ from $10^9$ cm$^{-3}$ to $10^{19}$ cm$^{-3}$ and compare the population of the density sensitive metastable states to the population of the ground state, as shown in Figure \ref{fig:occ}. The occupation values for Figure \ref{fig:occ} show the results for an optically thick wind, with a column density $\log{(N_{\rm H}/\mathrm{cm}^{-2})} = 24.11$ (solid lines). We repeat the same process but for a wind with a lower column density of $\log{(N_{\rm H}/{\rm cm}^2)} = 22$ (dashed lines). 

We see that, while the column density does affect the calculations for the occupation of these states, the primary driver is the wind density. These slight differences are caused by particular transition calculations and photon escape fractions in {\sc pion} being sensitive to the column density. Photons in lines with a large optical depth may scatter many times, enhancing the probability that they will be absorbed. This process is more favorable for low optical depth transitions, whose photons can escape directly or with fewer interactions. When the photons escape directly, the probability that they will interact with and be absorbed by the metastable states is reduced, thereby reducing the population.

If absorption by the ground state and at least one metastable state for a given ion are visible in a high-resolution X-ray spectrum, it is possible to compare their ratio to a plot like this and determine the density of the outflow. 
\begin{figure*}[t!h]
    \centering
    \includegraphics[width=\linewidth, trim = 0 0 0 0, clip]{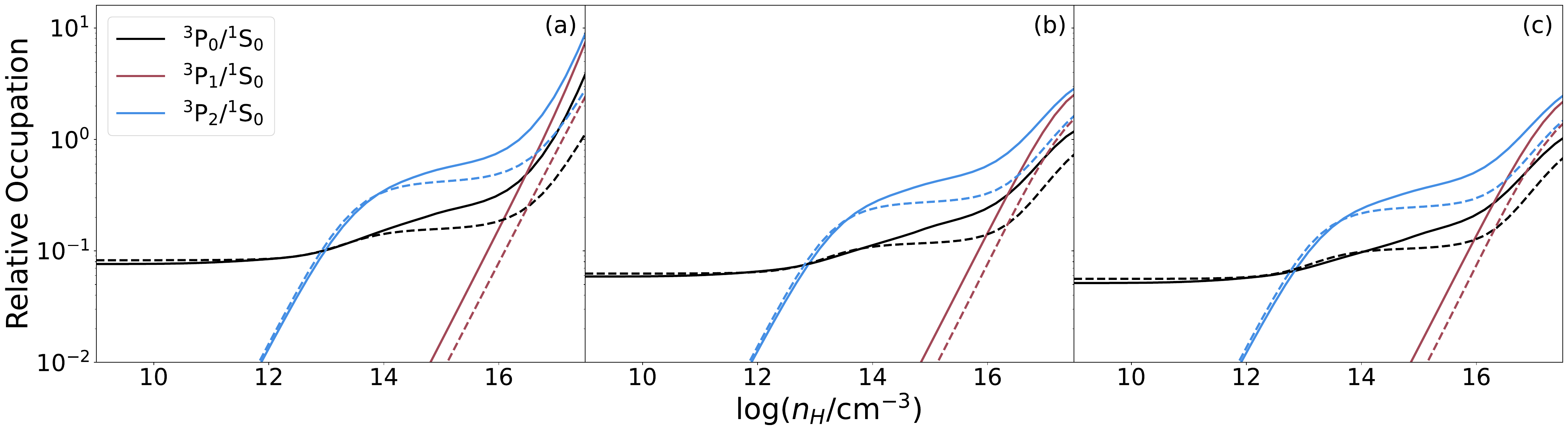}
    \caption{(a) The solid lines represent relative occupation ratios of the density sensitive metastable states of Fe{\sc XXIII} for an optically thick ($\log {(N_{\rm H}/{\rm cm}^2)} = 24.11$) wind using the LMXB1 SED. Dashed lines indicate the occupation with significantly lower column density ($\log{(N_{\rm H}/{\rm cm}^2)}=22$ cm$^{-2}$). We see that even with a severe reduction in $N_{\rm H}$, the overall shape remains, with differences apparent at densities $\log n_H \gtrsim 14$. This indicates that, while the column density can affect the population of metastable states, the primary driver is the wind density. (b) and (c) are the same, but for LMXB2 and LMXB3 respectively.  We cut the plots at a density of $\log (n_H/\mathrm{cm}^{-3}) \sim 17$, due to sudden breaks in the calculated occupations at that density. These are an indication that {\sc spex} has found a new solution for the occupations above those densities, and we do not consider these solutions to be physical. As mentioned in Table \ref{tab:lines} and Section \ref{sec:intro}}
    \label{fig:occ}
\end{figure*}

\section{Discussion}\label{sec:results}

After applying {\sc pion} to a range of ionizing SEDs, we see that the population of metastable states is dependent upon the precise nature of the source, as shown in Figures \ref{fig:conc} and \ref{fig:occ}. An LMXB with a hard, low flux continuum (such as LMXB2 and LMXB3) will produce different ratios of these states than that with a very high flux, thermally dominated continuum (like LMXB1). This is especially apparent in Figure \ref{fig:conc}, which shows that the concentration of the ions can change significantly depending on the SED. Here, LMXB3 looks distinct from LMXB1 and LMXB2 which are similar to one another. This is a result of the lack of thermal contribution in LMXB3. In order to accurately measure the density of the outflow, we consider a handful of other factors in this section. 

We also test the effect of the UV contribution in these different SEDs by applying an arbitrary low energy cut-off to LMXB2, removing emission below $\sim100$ eV. We find the population looks very similar to that of LMXB1, which also has reduced UV emission. The primary differences between LMXB1 (and therefore the artificially cutoff LMXB2) and the other SEDs seems to be a reduction in metastable populations, especially at high densities. When looking at the transitions below 100 eV, we find that many more of them de-populate the metastable states of Fe {\sc XXIII} than populate them, and at higher rates than the populating transitions.

\subsection{Fe K$\alpha$ Reflection} \label{subsec:refl}
Additional emission in the energy bands near the metastable transitions will contribute to the population of those states. Many LMXBs display a strong relativistically broadened Fe fluorescence feature near 6.4 keV, caused by the disk absorbing and re-processing seed photons from near the compact object \citep{fabian89}. The Gaussian component in the LMXB1 SED represents this emission, but it is not included in LMXB2 or LMXB3 (Table~\ref{tab:seds}). We test the impact of this component by comparing LMXB1 both with and without the reflection component. In both cases, we use the optically thick regime ($\log{N_{\rm H}} = 24.11$ cm$^{-2}$), while only changing the normalization of the Gaussian component to 0 in the SED. We find, however, that this yields  relative differences in the population ratios (like those seen in Figure \ref{fig:occ}) of order $\sim1\%$ at most densities.

\subsection{Feature Depth and Observability} \label{subsec:depth}
In order to determine the feasibility of observing the $^3P_1$ and $^3P_2$ metastable features, we estimate the optical depth at line center $\tau_0$ using:

\begin{equation}\label{eq:depth}
    \tau_0 = \sqrt{\frac{\pi}{2\sigma^2_v}}\frac{e^2\lambda}{m_ec}N_{\rm H}A_{Fe}C_i\sum_jf_jo_j
\end{equation}

where $e$ and $m_e$ represent the electron charge and mass, and $\lambda$ is the wavelength of the line centroid (adapted from equation 9.9 of \citealt{draine11}). The column density for the ion of interest is represented by $N_{\rm H} A_{Fe} C_i$, where $N_{\rm H}$ is the H column density, $A_{Fe}$ is the abundance ratio of Fe/H, and $C_i$ is the concentration, or fraction of Fe in the ionization state $i$. For transitions from atomic state $j$, the optical depth is proportional to the oscillator strength $f_j$ and $o_j$, the occupation fraction for that atomic energy level. We use $\tau_0$ rather than something like equivalent width, since it is a value more directly extracted from most line-fitting models. Regardless, in such a scenario where the velocity is kept constant across the entire grid, as it is here, the equivalent width and $\tau_0$ would be roughly proportional, yielding similar relative results. We choose to focus on the blended $^3P_{1,2}$ feature due to its close proximity to a contaminating Fe~{\sc XXIV} feature. In Section~\ref{subsec:contamination}, we will compare the strength of the contaminating Fe~{\sc XXIV} line to these features of interest.

Since the metastable state occupations depend strongly on both ionization and density, we use Equation \ref{eq:depth} to calculate the relative line strengths for a grid of values, varying $\log{\xi}$ from $2-4$ in 16 steps and $\log{(n_H/{\rm cm^{-3})}}$ from $9-19$ in 16 steps. However, after $\log{(n_H/{\rm cm^{-3})}} \approx 17$, {\sc pion} calculations became unable to find a continuous solution, so we cut our plots at that value. 
For this calculation, we employ the optically thick case of $\log{(N_{\rm H}/{\rm cm^{-2})}} = 24.11$ and with $\sigma_v = 100$ km/s, based off of the slow wind from XC25. We use $\lambda=1.874$ \AA\ as the central wavelength of the $^3P_1$ and $^3P_2$ features}. $C_i$ is taken from the {\sc spex} models using the {\sc icon} table. We use the solar system abundance $A_{Fe} = 2.95\times10^{-5}$ from \cite{lodders09}. The $f_j$ values are given in Table~\ref{tab:lines} and the $o_j$ values are taken from {\sc spex} using the {\sc pop} table. 
The results are shown in the heatmap of Figure \ref{fig:heatmap_depth}. While the choice of $N_{\rm H}$ will affect the abundance of each metastable state (Figure~\ref{fig:occ}), we are primarily interested in the distribution of optical depths for various values of $\xi$ and $n_H$. We see a distribution which tracks closely to our understanding of these density-sensitive features; namely, the depth increases as we approach the peak concentration of Fe {\sc XXIII} at $\log \xi \approx 2.75$, with higher optical depth at higher densities. 

The populations (and therefore the calculated depths of the features) are also dependent on the uncertainty in the atomic data, such as the central wavelength of the absorption feature. However, these uncertainties are low; the values from \cite{steinbrugge22} are quite precise, at $<0.1$ eV. The values from NIST are slightly higher, at around 0.5 eV. Even with the more conservative uncertainties, this results in a relative error of $<0.01\%$. With these precise values, we therefore consider the trends to be accurate.

\begin{figure*}[t!h]
    \centering
    \includegraphics[width=0.98\linewidth, trim = 0 0 0 0, clip]{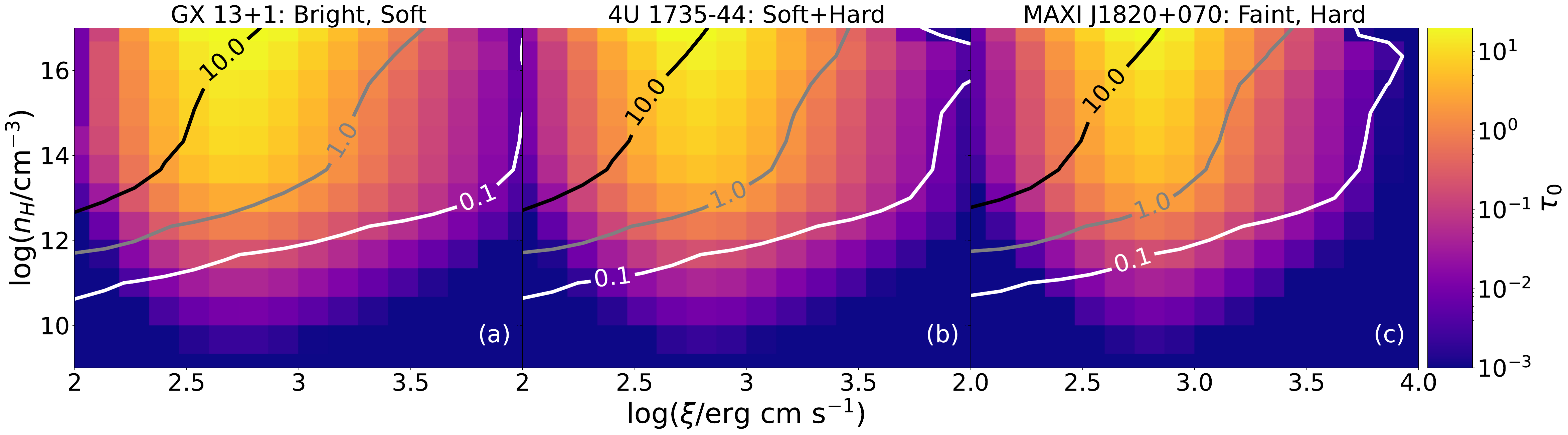}
    \caption{(a) For LMXB1, we display the optical depth at line center $\tau_0$ as calculated by Equation \ref{eq:depth} for the combined $^3P_1$ and $^3P_2$ feature of Fe {\sc XXIII}. The contours represent the ratio of $\tau_0$ for the features found at $\sim6.617$ keV for Fe {\sc XXIII} and {\sc XXIV} respectively (i.e., a value $> 1$ indicates the Fe {\sc XXIII} feature is stronger than the contaminating Fe {\sc XXIV}). We use black, grey, and white contours to represent a ratio of 10, 1.0, and 0.1, respectively. (b) and (c) are the same, but for LMXB2 and LMXB3 respectively. }
    \label{fig:heatmap_depth}
\end{figure*}

\subsection{Contamination from Fe {\sc XXIV}} \label{subsec:contamination}

Figure \ref{fig:conc} shows that that the concentration of Fe{\sc XXIV} and Fe{\sc XXIII} peak at similar values of $\xi$, indicating that within a relatively small range of ionization, Fe {\sc XXIV} becomes a significant source of contamination of density sensitive Fe {\sc XXIII} features. As shown in Table \ref{tab:lines}, these features correspond to density sensitive features of Fe {\sc XXIII}. 

We next account for contamination by Fe {\sc XXIV}. To determine the relative strengths of the Fe~{\sc XXIII} and Fe~{\sc XXIV} features of interest, we use Equation \ref{eq:depth}, dividing $\tau_0$ for the Fe {XXIII} feature arising from the combined $^3P_1$ and $^3P_2$ states by the value of $\tau_0$ calculated for the Fe {\sc XXIV} feature that contaminates it. For the Fe {\sc XXIV} feature, we use the oscillator strength $f = 0.0166$ \citep{mehdipour15b}. We assume the central energy of the Fe {\sc XXIII} and Fe {\sc XXIV} features are close enough that we treat their $\lambda$ values as equal when adapting from Equation \ref{eq:depth}. The same grid and assumptions used in Section \ref{subsec:depth} is used, as well as the same column density.

The contours in  Figure \ref{fig:heatmap_depth} represent the ratio of the optical depth of the $^3P_{1,2}$ Fe {\sc XXIII} feature to the contaminating Fe {\sc XXIV} feature. This shows that the regime in which the Fe {\sc XXIII} features are more prominent than those of the contaminating Fe {\sc XXIV} is approximately in the density range $\log{(n_H/{\rm cm^{-3)}}}\gtrsim14$ (with a maximum depth around  $\log{(n_H/{\rm cm^{-3)}}}\approx17$) and the ionization range $\log(\xi/{\rm erg~cm~s^{-1}})\sim2-3$. This range represents the low end of typical NS LMXB ionization parameters $\log\xi \sim 2.5-3.5$ erg cm s$^1$ \citep{cackett10, ludlam17, ludlam19a}. GX~13+1, the LMXB on which the LMXB1 SED was based, typically persists at the upper end of this range, with a $\log(\xi/{\rm erg~cm~s^{-1}})\approx4$ (\citealt{allen18}, XC25). While the spectrum of GX~13+1 is rich with wind absorption signatures, this implies that it will not likely be a feasible candidate for such an analysis. 

However, for other sources this method becomes more reasonable. For example, BH LMXBs, like MAXI J1820+070 (the source used to generate the LMXB3 SED), can in some cases have ionization parameters which extend to both higher and lower values than their NS LMXB counterparts, with $\log(\xi/{\rm erg~cm~s^{-1}}) \sim 1.8-6$ \citep{miller04, king14, diaztrigo13a, higginbottom17, connors21, marino21}. For this reason, we suggest that it is more likely that BH LMXBs will be an avenue for density diagnostics using Fe {\sc XXIII} metastable transitions than their NS counterparts.

Additionally, other nearby lines could be used to further disentangle the contamination from Fe XXIV. There are several other nearby transitions from the metastable states that could be used to better constrain the relative contributions of Fe XXIII and Fe XXIV. Similarly, the relative strengths of other Fe XXIV lines can be used to estimate the strength of the contaminating Fe XXIV feature. {The strong $q$ ($E_0=6.6622$ keV), $r$ ($E_0=6.6528$ keV), and $t$ ($E_0=6.6762$ keV) lines  \citep{rudolph13} can be used alongside their oscillator strengths in order to estimate the strength of the $u$ line, which is the contaminating feature in Table \ref{tab:lines}.

\section{Summary and Conclusions} \label{sec:conc}
The ability to estimate the density of an outflow in an LMXB disk is useful for understanding the geometry and energy budget of such systems. One such method for doing this is to use density-sensitive metastable absorption lines. In this paper, we explore the process and feasibility of using transitions of Fe {\sc XXIII} to estimate the density of a photoionized absorber in an LMXB system. We test the feasibility of this technique using a range of ionization parameters and wind densities for three distinct ionizing SEDs, and compare the strength of relevant features. 

We find that this technique is feasible for a range of ionization parameters and densities; in general, we find that the technique is possible for values $\log\xi\sim2-3$ erg cm$^{-1}$ s$^{-1}$ with an outflow density $\log{(n_H/{\rm cm^{-3})}}\gtrsim14$. Using Fe {\sc XXIII} lines as a density diagnostic is perhaps most feasible in BH LMXBs. In such cases, the ionization parameter and density required is in the range of previously measured disk winds. However, difficulties arise at higher ionizations, where Fe {\sc XXIV} contamination is an important factor.

It is perhaps possible to account for all of these obstacles, even outside of the limited regimes discussed above. By carefully modeling the ionizing SED (including contributions from UV photoexcitation) and using photoionization software like {\sc pion}, one could track the expected ratios of Fe {\sc XXIII} and {\sc XXIV}, and use that information to account for contamination. This requires high signal-to-noise and resolution, such as what is capable with XRISM Resolve, allowing for the small contributions of Fe {\sc XXIII} to be carefully measured and compared against the more significant Fe {\sc XXIV}. XRISM {\it Resolve} is uniquely capable of discerning these faint lines and estimating the density of LMXB outflows. Understanding the nature of disk winds and the information they can provide continues to be a fascinating challenge. We have shown that in certain regimes, it may be a useful metric in the era of high-resolution X-ray spectroscopy.

{\it Acknowledgments:} This work was made possible by support from NASA grants 80NSSC18K0978, 80NSSC20K0883, 80NSSC25K7064, 80NSSC20K0733, 80NSSC24K1148, 80NSSC24K1774, and 80NSSC23K0684. Support is also provided from STFC through grant ST/T000244/1. This work was supported by JSPS Core-to-Core Program, (grant number:JPJSCCA20220002). This research has made use of data and/or software provided by the High Energy Astrophysics Science Archive Research Center (HEASARC), which is a service of the Astrophysics Science Division at NASA/GSFC. Additional special thanks to Dr. Junjie Mao, Dr. Jelle Kaastra, and Dr. Missagh Mehdipour, who provided invaluable feedback and useful conversations surrounding this study. 

\bibliographystyle{aasjournal}
\bibliography{bibliography}

\end{document}